\documentstyle[prb,aps,epsfig,twocolumn]{revtex} 
 
\begin{document} 
\flushbottom 
\twocolumn[ 
\hsize\textwidth\columnwidth\hsize\csname @twocolumnfalse\endcsname

\title{Hybridization of localized and density modes for 
 the roton spectrum of superfluid ${^4}$He.} 
\author{N.~Gov and E.~Akkermans\\ Department of Physics, Technion, 32000 
Haifa, Israel} 
\maketitle 
\tightenlines 
\widetext 
\advance\leftskip by 57pt 
\advance\rightskip by 57pt

\begin{abstract} 
A new description is presented for the roton part of the energy spectrum of superfluid  
${^4}$He. It is based on the  
assumption that in addition to the Feynman 
 density excitations, there exist localized modes 
 which describe vortex-core elements. 
The energy spectrum which results from  
 the hybridization 
 of these two kinds of excitations  
is compared to the experimental data namely the  
structure factor and the scattering cross-section for the single quasiparticle 
 excitation.  
Another type of excitation interpreted as  
a vortex loop is obtained whose energy agrees both with Raman  
scattering and critical velocity experiments. 
\end{abstract} 
 
\vskip 0.2cm 
] 
 
\narrowtext 
\tightenlines 
 
\vspace{.2cm} 
 
In this letter we present a new theoretical description of the roton region 
of the excitation spectrum of superfluid ${^4}He$. In spite of the success 
obtained using the wavefunction proposed by Feynman ~\cite{feynman,nozieres}%
, a quantitative agreement with the experimental results has not yet been 
achieved. Subsequent refinements using more complicated variational 
wavefunctions were at the expenses of a simple interpretation as a pure 
density fluctuation \cite{reatto}. More recently, new experimental results ~%
\cite{stirling} obtained in high resolution neutron scattering did show 
puzzling features such as the persistence of the phonon peak up to 
temperatures higher than $T_\lambda $ together with the decrease of the 
sharp peaks associated respectively to the maxon and roton excitations. 
These features tend to indicate that the low momentum phonon mode is 
marginally related to superfluidity or even to the Bose-Einstein statistics, 
a point which was already noticed by Pines ~\cite{pines} who proposed to 
interpret this mode as a collective zero sound just like in $^3{He}$. In 
contrast, the decrease at the transition of the other modes suggests a 
connection with quantum exchange effects. 
 
In order to interpret these features we assume the existence of two kinds of 
excitations in the system. At large scales, there are delocalized density 
fluctuations while at scales of a few interatomic distances there exist 
localized excitations which may be associated with microscopic vortex-core 
elements. These two kinds of excitations are not independent but instead are 
hybridized in a way reminiscent of the case of excitons in dielectric 
crystals ~\cite{hopfield}. This analogy will guide us in order to build a 
phenomenological Hamiltonian. This phenomenological model will allow us to 
derive analytical expressions for measurable quantities and to provide an
 interpretation for the high-momentum roton region of the spectrum 
where the variational approache becomes more involved. 
 
The physical motivation leading to the assumption of the existence of localized excitations 
stems from the following remarks. In any attempt to reproduce quantitatively 
the spectrum using a variational approach \cite{reatto} more localized 
structure needs to be introduced into the excited-state wavefunction. In addition it 
is generally accepted that the superfluid supports quantized vortices with microscopic 
cores \cite{core}. In order to describe them, it is relevant to consider ~\cite{ceperley}
 exchange effects associated to localized and closed exchange rings of 
atoms. Therefore, we view the system as being a liquid at large distances 
while locally, on the scale of some interatomic distances, we consider 
localized quantum clusters playing the role of the vortex-core elements. Since 
the helium atoms are indistinguishable, we cannot split them into two 
classes (i.e. either embedded in a long range density mode or in a localized 
cluster) but we suppose instead that each atom retains both characteristics. 
As a consequence, these two kinds of excitations are coupled and the 
spectrum results from this hybridization. From a methodological point of 
view, our approach is close to that of Hopfield ~\cite{hopfield} who considered  
excitons in dielectric crystals as localized excitations whose 
interactions are mediated by the photons. Here, along the same line, the 
interaction between the localized excitations is mediated by the Feynman 
density fluctuations. 
 
To make those qualitative remarks more precise, we first consider the 
Hamiltonian ${H_0}=\sum_k\epsilon (k){{a_k}^{\dagger }}{a_k}$ describing the 
pure density modes obtained using the Feynman ansatz ~\cite{feynman} with 
the energy $\epsilon (k)={\frac{{\hbar ^2}{k^2}}{{2mS(k)}}}$ , where $m$ is the 
mass of the helium atoms and $S(k)$ the structure factor of the liquid. 
Then, we assume the localized excitations to be two-level systems of energy $%
\hbar {\omega _0}$, such that the corresponding Hamiltonian is ${H^0}_{loc}={%
\sum_k}\hbar {\omega _0}\left( {{b_k}^{\dagger }}{b_k}+{\frac 12}\right) $. 
The operators $a$ and $b$ obey bosonic commutation relations and 
commute between themselves, but for the $b$ 
operators the bosonic character is only approximate and holds in the limit of a low density of 
localized modes ~\cite{anderson}. To write down the part describing 
interactions between the two sets of excitations, we follow the approach of 
Hopfield and Anderson ~\cite{hopfield,anderson} and assume that the role of 
the density fluctuations is to induce an effective interaction between the 
localized modes which within the dipolar approximation allows to 
write the effective Hamiltonian  
\begin{eqnarray} 
{H_{loc}} &=&{\sum_k}(\hbar {\omega _0}+X(k))\left( {{b_k}^{\dagger }}{b_k}+{%
\frac 12}\right)  \nonumber \\ 
&&\ \ +{\sum_k}X(k)\left( {{b_k}^{\dagger }}{b_{-k}^{\dagger }}%
+b_kb_{-k}\right) 
\end{eqnarray} 
where $X(k)$ is a (real and negative) matrix element which depends on the 
microscopic details of the dipolar interaction.
The dipolar approximation is the simplest limit for which the resulting Hamiltonian can 
be analytically worked out. Moreover, it corresponds to the physical picture of 
the localized-modes as vortex-core elements. Such microscopic line-elements 
may be treated as local dipolar defects with respect to the surrounding fluid.
 As we shall see later, in this limit, the knowledge of the  
exact expression of $X(k)$ is not necessary. 
The coupling between the phonons and the localized modes is described 
by the Hamiltonian ~\cite{hopfield}  
\begin{eqnarray}
H_c = {\sum_k} \left( \lambda (k,{\omega_0}) {b_k} +\mu
(k,{\omega_0}){a_k}
\right)
({{a_k}^\dagger}+{a_{-k}}) + h.c.
\end{eqnarray}
where the two functions
 $\lambda$ and $\mu$ are given by
$\lambda (k) = i {\hbar\omega_0} (-{3X(k) \over {2 \epsilon (k)}} {)^{1
\over 2}}$ and
$\mu(k) = - {\hbar\omega_0} {{3 X(k)} \over {2 \epsilon (k)}}$.
${H_{loc}}$ is diagonalized using the Bogoliubov transformation
${\beta_k} = u(k) b_{k} + v(k) {b^\dagger}_{-k}$, which the resulting spectrum
\begin{equation} 
E=\sqrt{{\hbar \omega _0}({{\hbar \omega _0}+2X(k))}} 
\end{equation} 
The two functions $u(k)$ and $v(k)$ are given by
${u^2}(k)={1 \over 2} ({ {{\hbar\omega_0} +X(k)} \over {E(k)}} +1)$ and
${v^2}(k)={1 \over 2} ({ {{\hbar\omega_0} +X(k)} \over {E(k)}} -1)$.
The Hamiltonian $H_0+{H^0}_{loc}+H_c$ is quadratic and is diagonalized
by
means of the
canonical transformation $\alpha_{k} = A a_{k}+ B b_{k}+ C
{a^\dagger}_{-k} + D
{b^\dagger}_{-k}$ and
${\tilde{\alpha}}_{k} = B a_{k}+ A b_{k} + D {a^\dagger}_{-k} + C
{b^\dagger}_{-
k}$.
The corresponding dispersion
relation is
\begin{equation}
{{\epsilon^2}(k) \over {{E^2}(k)}}= 1- {6 \over{\hbar\omega_0}}{{X(k)
\over {1 - ({E(k) \over  {\hbar\omega_0}})^2}}}
\end{equation}
We first notice that taking the coupling $X(k)$ between the two sets of 
modes to zero, we obtain, as expected, the two solutions $E(k)=\epsilon (k)$ 
describing a pure density mode and $E={\hbar \omega _0}$ for the localized 
two-levels. A non zero coupling hybridizes these two sets of excitations. We 
take for the energy $\hbar \omega _0$ the highest value of the 
phonon-roton spectrum i.e. where it terminates, namely ${\hbar \omega _0}%
=2\Delta $ where the energy $\Delta $ corresponds to the roton minimum. 
New experimental results at higher values of $k$ confirm this behaviour at the 
termination point \cite{glyde}.
 To solve for the dispersion relation and to obtain the phonon part of the spectrum, 
we use (3) into (4). This gives $E(k)={\frac 12}\epsilon (k)$ i.e. an 
expression independent of the matrix element $X(k)$. Using now this latter 
expression into (4), we obtain the other branch describing the hybridized 
local modes at $E=2{\hbar \omega _0}=4\Delta $ which is as well independent 
of $X(k)$. We emphasize again that this is a consequence of the particular choice 
of the dipolar approximation \cite{hopfield} which is manifested in the expressions for the 
coupling functions $\lambda $ and $\mu $. As a 
result of the hybridization, the energy spectrum of the density fluctuations 
is shifted by a factor two towards lower energies relatively to the Feynman 
ansatz, and the localized modes still have a constant energy although twice the original 
value. 
 
To compare our results with the experimental data we shall consider first 
two independent sets of results namely the measurements of the energy 
spectrum and of the structure factor $S(k)$. We obtain from (4) the relation  
\begin{equation} 
E(k) = {\frac{{\hbar^2}{k^2} }{{4m S(k)}}} 
\end{equation} 
for the lower branch which we compare to the experimental results obtained 
at two different pressures both for $E(k)$ \cite{cowley,donely} and $S(k)$  
\cite{sven,hen,hall} (Figs.1 and 2). The main discrepancy is obtained in the 
low momentum region ($k \leq 1 {\AA}^{-1}$), i.e. below the maxon momentum. 
As $k \rightarrow 0$ the experimental data is close to the Feynman result i.e.
 twice the value we obtain in (5). This signals a failure of the 
dipolar approximation for long wavelengths. Our description becomes meaningless 
when the phonon-roton branch terminates since the Feynman spectrum 
approaches that of a free particle. Around the roton minimum, and over a 
large range of momentum the result (5) provides a good fit to the 
experimental results. Feynman himself in his original paper \cite{feynman}
 noticed the factor 2 
discrepancy between his ansatz and the experimental results around 
the roton minimum. Here we deduce this 
factor from an exact analytical solution. This interpretation of the roton 
excitation as resulting from the dipolar hybridization of two separate 
excitations is to be compared with the approach of Glyde and Griffin ~\cite 
{Glyde and Griffin,griffin} which uses a dielectric formalism to 
describe hybridized phonons and free-particle excitations. 
 
The neutron scattering intensity is a direct measure of the density 
fluctuations in the liquid and is obtained from the dynamical structure factor  
$S(k, \omega)$. It is accepted since the work of Miller, Pines and 
Nozieres ~\cite{mpn} that we can split $S(k, \omega)$ into two parts, $S(k, 
\omega)= N Z(k) \delta (\hbar\omega- {\epsilon_k}) + {S^{(1)}}(k, \omega)$ 
where the first term accounts for single quasi-particle excitations while 
the second describes multiparticle excitations. This separation is quite 
easy to justify at low temperature and low momentum, typically for $k \leq 
0.5 {\AA}^{-1}$ where we obtain $Z(k)= S(k)$, which results also from the 
Feynman theory. The comparison with the experimental data ~\cite{cowley} 
shows that although it works in the low momentum regime mentionned above, it 
fails to describe the behavior at higher momentum, except perhaps for the 
position of the maximum. In contrast to this, we do not consider here a precise 
decomposition of the structure factor. Since the density fluctuations 
are hybridized with the localized modes, we expect the differential 
cross-section $Z(k)$ for the excitation of a single quasi-particle to be 
proportional to $\langle {{b_k}^\dagger}{b_{-k}^\dagger} \rangle = {u_k}{v_k} 
$, where the expectation value is calculated in the ground state of the 
Bogoliubov pairs. Then, $Z(k) = 4 \pi {k^2} {I_0} {u_k}{v_k}$ where $I_0$ is 
a normalization constant. Using our previous expression for $u_k$ and $v_k$ 
we obtain ${u_k}{v_k} = {\frac{1 }{2}} {\frac{{|X(k)|} }{{E(k)}}}$ which 
together with (3) and (5) gives  
\begin{equation} 
Z(k) = \pi {k^2} {I_0}{\frac{{\hbar {\omega_0}} }{{E(k)}}} \left| \left( {%
\frac{{E(k)} }{{\hbar {\omega_0}}}} \right)^2 -1 \right| 
\end{equation} 
From the two independent measures of $E(k)$ and $S(k)$ we obtain using (6) a 
theoretical expression of the differential cross-section which fits well the 
experimental results as shown in Fig.3. In particlular from (6) it follows 
that the intensity of the single quasi-particle branch vanishes  
when $E(k) \rightarrow 2\Delta$ which is a new result. 
Moreover, in the low momentum limit, we recover the expected proportionality between $%
S(k)$ and $Z(k)$. 
 
We found a second branch of excitations at the constant energy $E=4\Delta $. 
It describes localized excitations of energies twice the bare vortex core 
energy. It is suggestive to interpret this mode as a single vortex loop 
whose radius can be calculated using a Feynman-type formula for the energy 
of the circulating current of a vortex-loop ~\cite{nozieres}  
\begin{equation} 
E_{vortex-loop}=2{\pi ^2}\rho {\frac{{\hbar ^2R}}{{m^2}}}ln\left( {\frac{{R}%
}{{a}}}\right) 
\end{equation} 
where at $T=0$, the density of the superfluid is $\rho ={\rho _s}$ 
and $a$ is the core size equal to the atomic radius \cite{core} namely $a\simeq 1.4\AA $.
 The radius $R$ is obtained taking $E_{vortex}=4\Delta =34.4K$. 
This gives $R\simeq 5.0\AA $ which is the expected size for the smallest 
vortex-loop. A further experimental evidence in favour of this 
interpretation is provided by critical velocity experiments ~\cite{varoquaux}%
. In phase-slippage studies of the critical velocity through an orifice, the 
critical velocity is driven by the thermal nucleation of vortex-loops. The 
corresponding energy $E_v$ is determined by the nucleation rate $\Gamma $ 
which, using the Arrhenius law  $\Gamma ={\Gamma _0}exp\left( {\frac{{-{E_v}}}{{%
{k_B}T}}}\right) $, is found ~\cite{varoquaux} to be ${E_v}\simeq 33\pm 5K$, 
which is indeed very close to our result $4\Delta =34.4K$. Finally, the 
upper critical velocity $v_c$ may be estimated as the velocity of the 
vortex-loop itself ~\cite{nozieres} namely ${v_c}={\frac{{\hbar }}{{2mR}}}%
ln\left( {\frac Ra}\right) \simeq 20m/s$, a value close to the largest 
measured critical velocity ~\cite{varoquaux}. 
 
Another relevant set of experiments we consider is provided by the 
Raman scattering around $k=0$. It has been found that besides a peak at $E= 
2 \Delta$, there is an additional contribution at $E = 4 \Delta$ ~\cite 
{ohbayashi} which we associate to the vortex-loop. The peak at $2 \Delta$ is 
usually interpreted as a two-roton excitation and therefore the additional 
contribution is viewed as a four-roton excitation. This interpretation 
suffers nevertheless from the fact that there is no three-roton peak. In our 
model, the lowest excitation energy of a vortex-core is given by $2 \Delta$ 
and not by $\Delta$ so that we do not expect any contribution at $3 \Delta$. 
The peak at $2 \Delta$ which does not appear in the hybridized spectrum will 
be discussed elsewhere ~\cite{gov}. 
 
In conclusion, we have shown that the roton region of the excitation 
spectrum of superfluid ${^4}He$ can be described quantitatively by assuming 
the existence of two kinds of excitations. One is provided by the 
delocalized density fluctuations given by the Feynman variational ansatz. In 
contrast, the second set of excitations describes the short range order in 
the system around a microscopic vortex-core element. We have assumed that 
these excitations are coupled in a way reminiscent from the case of excitons 
in dielectric systems. By writing a phenomenological Hamiltonian together 
with the dipolar approximation for the coupling, we obtained an excitation 
spectrum which involves two sets of hybridized modes. One corresponds to
 phonons shifted by a factor of two towards lower energies 
relatively to the Feynman result. The second set is interpreted as the 
intrinsic quantized vortex-loop modes of the superfluid. This picture 
provides analytical expressions which describe quantitatively a broad range 
of experimental results. In addition, it may help understanding the 
persistence of the phonon peak above $T_\lambda $ which appears to be 
shifted upwards in energy together with a sharp drop of both the maxon and 
roton peaks. Since the localized excitations depend on the superfluid order, 
they will vanish at the transition to the normal state, while the Feynman 
density modes will remain unaffected and only shifted upwards in energy in 
the abscence of dipolar coupling $X(k)$. Our model provides a new picture of 
the nature of the roton excitation and may allow for a unified 
treatment of the phonon-roton elementary excitations and of the quantized 
vortex-loops excitations which are both unique to the superfluid phase \cite 
{putterman}. 
 
{\bf Acknowledgement} This work is supported in part by a grant from the 
Israel Academy of Sciences and by the fund for promotion of research at the 
Technion.

\begin{figure}
{\hspace*{-0.2cm}\psfig{figure=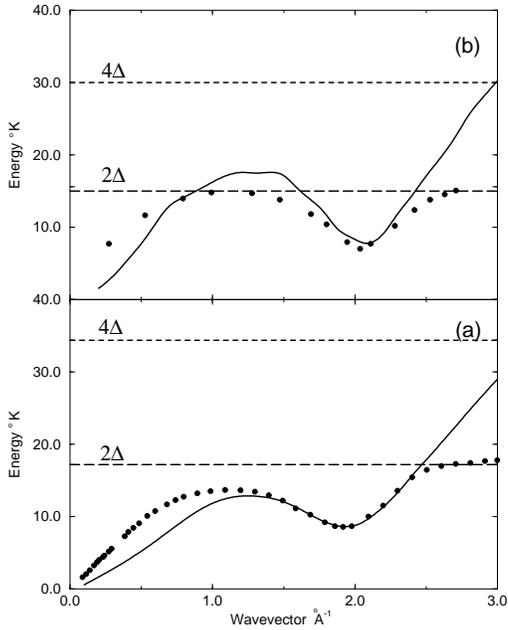,height=9cm,angle=0}}
\caption{Comparison between the experimental energy spectrum
[10,11] (points) and
the theoretical expression (5)(solid line) where the structure factor
$S(k)$ is
obtained from independent measurements [12,13]. (a) and (b)
correspond respectively to
the saturation vapor pressure and to P=24 atm. The dashed line at
$4\Delta$ indicates the position of the branch of the vortex-loop
excitations.}
\end{figure}
\begin{figure}
{\hspace*{-0.2cm}\psfig{figure=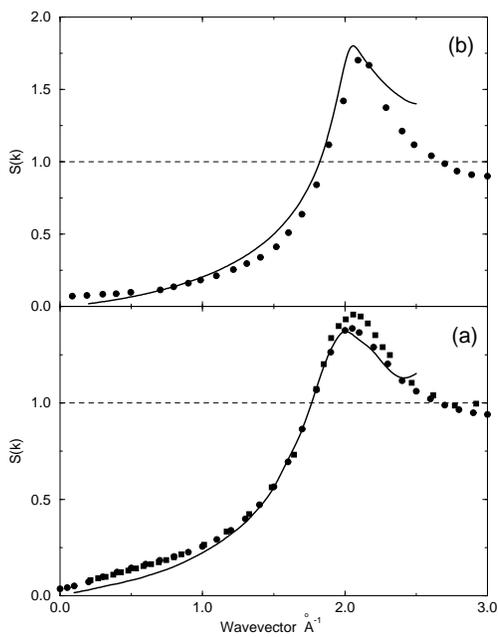,height=9cm,angle=0}}
\caption{Comparison between the experimental structure factor
$S(k)$ (points) and the expression (5) (solid line) for the
same two pressures as in fig.(1) where the energy $E(k)$ is obtained
from independent measurements [10,11]. (a) circles [12],
squares [14], (b) circles [13].}
\end{figure}
\begin{figure}
{\hspace*{-0.2cm}\psfig{figure=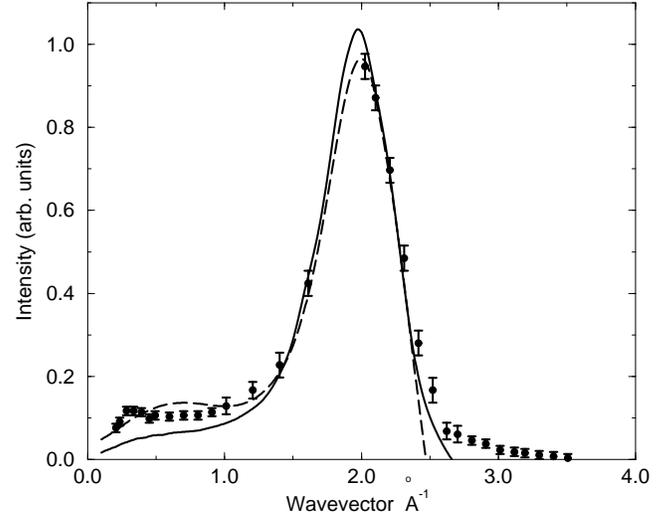,height=8cm,angle=0}}
\caption{Comparison between the experimental scattering cross-section
[10] $Z(k)$ of single quasi-particle excitations (points) at 1.1 K and
the theoretical expression (6). The two curves are obtained using
respectively in
(6) the experimental results for $S(k)$ [12] (dashed line) and for
the energy $E(k)$ [11] (solid line).}
\end{figure}
\end{document}